\documentclass[%
 reprint,
superscriptaddress,
 amsmath,amssymb,
 aps,floatfix
]{revtex4-1}

\newcommand{\neb}{$\bar{\nu}_{\rm{e}} \; $}

\usepackage{graphicx}
\usepackage{dcolumn}
\usepackage{bm}
\usepackage{amsmath}%
\usepackage{amssymb}%
\usepackage{subfigure} 
\usepackage[pagewise,mathlines]{lineno}

\begin{document}

\title{Background-independent measurement of $\theta_{13}$ in Double Chooz}

\newcommand{\Aachen}{III. Physikalisches Institut, RWTH Aachen 
University, 52056 Aachen, Germany}
\newcommand{\Alabama}{Department of Physics and Astronomy, University of 
Alabama, Tuscaloosa, Alabama 35487, USA}
\newcommand{\Argonne}{Argonne National Laboratory, Argonne, Illinois 
60439, USA}
\newcommand{\APC}{APC, AstroParticule et Cosmologie, Universit\'{e} Paris 
Diderot, CNRS/IN2P3, CEA/IRFU, Observatoire de Paris, Sorbonne Paris 
Cit\'{e}, 75205 Paris Cedex 13, France}
\newcommand{\CBPF}{Centro Brasileiro de Pesquisas F\'{i}sicas, Rio de 
Janeiro, RJ, cep 22290-180, Brazil}
\newcommand{\Chicago}{The Enrico Fermi Institute, The University of 
Chicago, Chicago, IL 60637, USA}
\newcommand{\CIEMAT}{Centro de Investigaciones Energ\'{e}ticas, 
Medioambientales y Tecnol\'{o}gicas, CIEMAT, E-28040, Madrid, Spain}
\newcommand{\Columbia}{Columbia University; New York, NY 10027, USA}
\newcommand{\Davis}{University of California, Davis, CA-95616-8677, USA}
\newcommand{\Drexel}{Physics Department, Drexel University, Philadelphia, 
Pennsylvania 19104, USA}
\newcommand{\Hamburg}{Institut f\"{u}r Experimentalphysik, 
Universit\"{a}t Hamburg, 22761 Hamburg, Germany}
\newcommand{\Hiroshima}{Hiroshima Institute of Technology, Hiroshima, 
731-5193, Japan}
\newcommand{\IIT}{Department of Physics, Illinois Institute of 
Technology, Chicago, Illinois 60616, USA}
\newcommand{\INR}{Institute of Nuclear Research of the Russian Academy 
of Science, Russia}
\newcommand{\CEA}{Commissariat \`{a} l'Energie Atomique et aux Energies 
Alternatives, Centre de Saclay, IRFU, 91191 Gif-sur-Yvette, France}
\newcommand{\Kansas}{Department of Physics, Kansas State University, 
Manhattan, Kansas 66506, USA}
\newcommand{\Kobe}{Department of Physics, Kobe University, Kobe, 
657-8501, Japan}
\newcommand{\Kurchatov}{NRC Kurchatov Institute, 123182 Moscow, Russia}
\newcommand{\MIT}{Massachusetts Institute of Technology; Cambridge, MA 
02139, USA}
\newcommand{\MaxPlanck}{Max-Planck-Institut f\"{u}r Kernphysik, 69117 
Heidelberg, Germany}
\newcommand{\Niigata}{Department of Physics, Niigata University, Niigata, 
950-2181, Japan}
\newcommand{\NotreDame}{University of Notre Dame, Notre Dame, IN 46556-
5670, USA}
\newcommand{\IPHC}{IPHC, Universit\'{e} de Strasbourg, CNRS/IN2P3, F-
67037 Strasbourg, France}
\newcommand{\SUBATECH}{SUBATECH, CNRS/IN2P3, Universit\'{e} de Nantes, 
Ecole des Mines de Nantes, F-44307 Nantes, France}
\newcommand{\Sussex}{Department of Physics and Astronomy, University of 
Sussex, Falmer, Brighton BN1 9QH, United Kingdom}
\newcommand{\Tennessee}{Department of Physics and Astronomy, University 
of Tennessee, Knoxville, Tennessee 37996, USA}
\newcommand{\TohokuUni}{Research Center for Neutrino Science, Tohoku 
University, Sendai 980-8578, Japan}
\newcommand{\TohokuGakuin}{Tohoku Gakuin University, Sendai, 981-3193, 
Japan}
\newcommand{\TokyoInst}{Department of Physics, Tokyo Institute of 
Technology, Tokyo, 152-8551, Japan  }
\newcommand{\TokyoMet}{Department of Physics, Tokyo Metropolitan 
University, Tokyo, 192-0397, Japan}
\newcommand{\Muenchen}{Physik Department, Technische Universit\"{a}t 
M\"{u}nchen, 85747 Garching, Germany}
\newcommand{\Tubingen}{Kepler Center for Astro and Particle Physics, 
Universit\"{a}t T\"{u}bingen, 72076 T\"{u}bingen, Germany}
\newcommand{\UFABC}{Universidade Federal do ABC, UFABC, Sa\~o Paulo, Santo 
Andr\'{e}, SP, Brazil}
\newcommand{\UNICAMP}{Universidade Estadual de Campinas-UNICAMP, 
Campinas, SP, Brazil}
\newcommand{\Aviette}{Laboratoire Neutrino de Champagne Ardenne, domaine 
d'Aviette, 08600 Rancennes, France}
\newcommand{\vtech}{Center for Neutrino Physics, 
Virginia Tech, Blacksburg, VA}
\newcommand{\deceased}{Deceased.}

\affiliation{\Aachen}
\affiliation{\Alabama}
\affiliation{\Argonne}
\affiliation{\APC}
\affiliation{\CBPF}
\affiliation{\Chicago}
\affiliation{\CIEMAT}
\affiliation{\Columbia}
\affiliation{\Davis}
\affiliation{\Drexel}
\affiliation{\Hiroshima}
\affiliation{\IIT}
\affiliation{\INR}
\affiliation{\CEA}
\affiliation{\Kansas}
\affiliation{\Kobe}
\affiliation{\Kurchatov}
\affiliation{\MIT}
\affiliation{\MaxPlanck}
\affiliation{\Niigata}
\affiliation{\NotreDame}
\affiliation{\IPHC}
\affiliation{\SUBATECH}
\affiliation{\Muenchen}
\affiliation{\Tennessee}
\affiliation{\TohokuUni}
\affiliation{\TohokuGakuin}
\affiliation{\TokyoInst}
\affiliation{\TokyoMet}
\affiliation{\Tubingen}
\affiliation{\UFABC}
\affiliation{\UNICAMP}
\affiliation{\vtech}

\author{Y.~Abe}
\affiliation{\TokyoInst}

\author{J.C.~dos Anjos}
\affiliation{\CBPF}

\author{J.C.~Barriere}
\affiliation{\CEA}

\author{E.~Baussan}
\affiliation{\IPHC}

\author{I.~Bekman}
\affiliation{\Aachen}

\author{M.~Bergevin}
\affiliation{\Davis}

\author{T.J.C.~Bezerra}
\affiliation{\TohokuUni}

\author{L.~Bezrukov}
\affiliation{\INR}

\author{E.~Blucher}
\affiliation{\Chicago}

\author{C.~Buck}
\affiliation{\MaxPlanck}

% ...

\collaboration{Double Chooz Collaboration}
\date{\today}
             
\begin{abstract}

The oscillation results published by the Double Chooz collaboration in 2011 and 2012 rely on background models substantiated by reactor-on data. In this analysis, we present a background-model-independent measurement of the mixing angle $\theta_{13}$ by including 7.53 days of reactor-off data. A global fit of the observed antineurino rates for different reactor power conditions is performed, yielding a measurement of both $\theta_{13}$ and the total background rate. The results on the mixing angle are improved significantly by including the reactor-off data in the fit, as it provides a direct measurement of the total background rate. This reactor rate modulation analysis considers antineutrino candidates with neutron captures on both Gd and H, whose combination yields $\sin^2(2\theta_{13})=$ 0.102 $\pm$ 0.028(stat.) $\pm$ 0.033(syst.). The results presented in this study are fully consistent with the ones already published by Double Chooz, achieving a competitive precision. They provide, for the first time, a determination of $\theta_{13}$ that does not depend on a background model.
\end{abstract}

\maketitle

\section{Introduction}

Recently, three reactor neutrino experiments, Double Chooz \cite{dcngd}, Daya Bay \cite{db} and RENO, \cite{reno} have successfully determined the leptonic mixing angle $\theta_{13}$ to be clearly non-zero. These disappearance experiments are sensitive to the oscillation amplitude and have measured $\sin^2(2 \theta_{13})$ to be $\sim 0.1$. They identify reactor antineutrinos via the inverse beta decay (IBD) reaction ${\bar \nu}_e p \to  e^+  n$ and use a coincidence between the prompt positron and the delayed neutron capture signals in order to separate antineurinos from background events.
However, correlated events due to fast neutrons, stopping muons and cosmogenic generated radio-nuclides form a dangerous background in experiments with shallow overburden.
So far, all published results are based on background models, which are derived from data taken during reactor-on periods using certain assumptions about the origin of correlated background events. 
This procedure contributes, along with detection efficiency and reactor source errors, to the total systematic uncertainty.
In this paper, we present a first measurement of  $\theta_{13}$ which is free from background assumptions. 

Of the three experiments, Double Chooz is the only one to be exposed to only two reactors. The total antineurino flux therefore changes significantly during reactor maintenance periods when one of the two reactor cores is not functioning. At certain times both cores at Chooz were turned off simultaneously, providing the unique opportunity to determine the background in a model independent way. In this paper we present an analysis of the Double Chooz data in which the background rate and the oscillation amplitude are determined simultaneously by analyzing the \neb candidate rates for different reactor conditions ranging from zero to full thermal power. The background rate has been proven to be constant in time \cite{dcoffoff}, thus being the same in all the considered reactor periods.
We restrict our analysis to rate measurements only.
In order to identify antineurino events via the inverse beta decay, 
we use both neutron captures on Gd and on H.
Finally, we present a combined Gd- and H-analysis and compare our final result with the published ones that rely on the energy spectrum information.
This analysis is also useful as a direct test of the background model used for the Double Chooz oscillation analysis.
We will show that our background rate determination is in full agreement with the prediction derived from our background model.

\section{Reactor Rate Modulation analysis}

In order to measure the mixing angle $\theta_{13}$ by means of reactor neutrino experiments, the observed rate of \neb candidates ($R^{\rm{obs}}$) is compared with the expected one ($R^{\rm{exp}}$). As Double Chooz data have been taken for different reactor thermal power ($P_{\rm{th}}$) conditions, this comparison can be done for different expected averaged rates, in a Reactor Rate Modulation (RRM) analysis. In particular, there are three well defined reactor configurations: 1) the two reactors are on (2-On data), 2) one of the reactors is off (1-Off reactor data), and 3) both reactors are off (2-Off reactor data). For the 1-Off and 2-Off reactor data, the expected antineurino rate takes into account the residual neutrinos ($ R^{\operatorname{r-\nu}} $) generated after the reactors are turned off as $\beta$ decays keep taking place. While the antineurino flux generated during reactor operation is computed as described in \cite{dcngd},  the rate of residual antineurinos is estimated as described in \cite{dcoffoff}.

From the comparison between $R^{\rm{exp}}$ and $R^{\rm{obs}}$ at different reactor powers both the value of $\theta_{13}$ and the total background rate $B$ can be derived. The correlation of the expected and observed rates follows a linear model parametrized by $\sin^2(2\theta_{13})$ and $B$:

\begin{equation}
\label{eq:model} 
R^{\rm{obs}}=B+R^{\rm{exp}}=B+\left(1-\sin^2(2\theta_{13})\eta_{\rm{osc}}\right)R^{\rm{\nu}},
\end{equation}

\noindent where $R^{\rm{\nu}}$ is the expected rate of actual antineurinos in absence of oscillation and $\eta_{\rm{osc}}$ is the average disappearance coefficient, $\langle\sin^2(\Delta m^2 L/4E)\rangle$. This coefficient is computed by means of simulations for each one of the data points as the integration of the normalized antineurino energy ($E$) spectrum multiplied by the oscillation effect driven by $\Delta m^2$ (taken from \cite{minos}) and the distance $L$ between the reactor cores and the detector. The average $\eta_{\rm{osc}}$ value corresponding to the full data sample is computed to be 0.55. Fitting the data to the above model provides a direct measurement of the mixing angle and the total background rate. In previous Double Chooz publications \cite{dcngd,dcnh}, the rates and the energy spectra of the three dominant background sources (fast neutrons, stopping muons and cosmogenic isotope $\beta$-$n$ decays) were estimated from reactor-on data, therefore building a background model that was fitted along with the mixing angle. In contrast, the RRM analysis extracts the total background rate from data in a model-independent and inclusive way, where all background sources (even possible unknown ones) are accounted for. The accuracy and precision on the fitted value of $B$, as well as on $\theta_{13}$, relies mostly on the 2-Off reactor data, as this sample provides a powerful lever arm for the fit. As the accidental background in the observed rate is known to 0.2\% by means of the off-time coincidences, the RRM analysis is performed with accidental-subtracted candidate samples. Therefore, hereafter the total background $B$ refers to all background sources except the accidental one. The RRM oscillation analysis can be performed separately with the \neb candidate samples obtained with neutron captures on Gd (n-Gd) and H (n-H), as well as with a combination of these. 

\begin{figure}
\begin{center}
\includegraphics[scale=0.4]{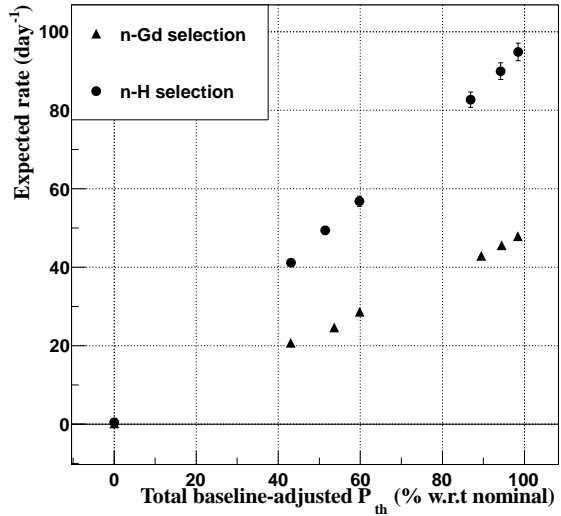}
\caption{ Expected unoscillated antineurino event rate as a function of the total baseline-adjusted thermal power ($P^{*}_{\rm{th}}=\sum^{\rm{N_r}}_{i} P^i_{\rm{th}}/L_i^{\rm{2}}$), for the n-Gd and n-H analyses. $P^{*}_{\rm{th}}$ is presented in percentage of the nominal power.
\label{fig:expvspth}}
\end{center}
\end{figure}

In the current analysis, the data sample in \cite{dcngd,dcnh} is used along with an extra 2-Off sample collected in 2012 \cite{dcoffoff}, which increases the total 2-Off run time to 7.53 days. Within the corresponding total live time of 233.93 days (246.4 days), 8257 (36883) candidates (including accidental background) were found according to the n-Gd (n-H) selection, 8 (599) of which were observed during the 2-Off period. The number of antineutrino events expected to be observed in the reactor-on periods was 8440 (17690). During the 1-Off period, the number of predicted residual events is 11.2 (28.7), while within the 2-Off period, 1.4 (3.7) residual events are expected in the n-Gd (n-H) selection. The data are distributed in 7 bins of $P_{\rm{th}}$, corresponding to two different sets of bins of $R^{\rm{exp}}$ for the n-Gd and n-H \neb candidate samples. The binning used for this analysis is shown in Fig.~\ref{fig:expvspth}, where the expected rates are presented as a function of the total baseline-adjusted thermal power, $P^{*}_{\rm{th}}=\sum^{\rm{N_r}}_i P^i_{\rm{th}}/L_{i}^{\rm{2}}$, where $N_{\rm{r}}$=2 is the number of reactors and  $L^{i}$ is the distance between the detector and reactor $i$. The error bars in the expected rates (not visible for all data points) account for the systematic errors.

\section{Systematic uncertainties}

There are three sources of systematics to be accounted for in the RRM analysis: 1) detection efficiency ($\sigma_{\rm{d}}$), 2) residual \neb prediction in reactor-off data ($\sigma_{\rm{\nu}}$), and 3) \neb prediction in reactor-on data ($\sigma_{\rm{r}}$). The detection efficiency systematics in n-Gd (n-H) \neb sample  are listed in \cite{dcngd} (\cite{dcnh}), from which the total uncertainty $\sigma_{\rm{d}}$ is derived to be 1.01\% (1.57\%). The uncertainty in the rate of residual antineurinos has been computed with core evolution simulations as described in \cite{dcoffoff} for the 1-Off and 2-Off reactor periods: a $\sigma_{\rm{\nu}}$=30\% error is assigned to $R^{\operatorname{r-\nu}}$. Finally, a dedicated study has been performed in order to estimate $\sigma_{\rm{r}}$ as a function of the thermal power.

\begin{figure}
\begin{center}
\includegraphics[scale=0.4]{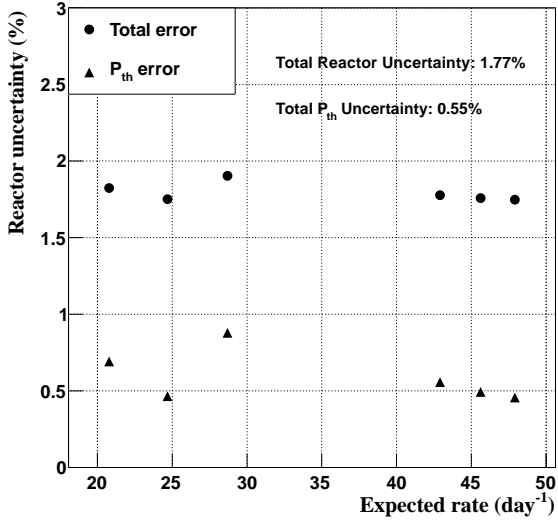}
\caption{ Uncertainty in the n-Gd \neb expected rate for reactor-on data. Triangles show the rate error due to $P_{\rm{th}}$ uncertainty, while circles stand for the total rate error accounting for all reactor-related systematics sources, as described and estimated in \cite{dcngd}.
\label{fig:reactsys}}
\end{center}
\end{figure}

To a good approximation, all sources of reactor-related systematics are independent of $P_{\rm{th}}$, with the exception of the uncertainty on $P_{\rm{th}}$ itself, $\sigma_{\rm{P}}$. This fractional error is 0.5\% \cite{dcngd} when the reactors are running at full power, but it increases as $P_{\rm{th}}$ decreases.  In \cite{dcngd, dcnh}, $\sigma_{\rm{P}}$ is assumed to be 0.5\% for all data. This is a very good approximation when one integrates all the data taking samples, and consequently all reactor operation conditions, as more than 90\% of the data are taken at full reactor power. However, this is not a valid approximation in the current analysis as it relies on separating the data according to different reactor powers. In order to compute $\sigma_{\rm{P}}$ for different $P_{\rm{th}}$, an empirical model is fitted to a sample of measurements provided by EdF (the company operating the Chooz nuclear plant). An effective absolute uncertainty of about 35 MW is derived from the fit, being the dominant component of the model. This absolute error translates into a $1/P_{th}$ dependence of the relative power uncertainty, which is used to compute the errors in $R^{exp}$. The resulting errors (both from $P_{\rm{th}}$ only and from all reactor sytematics sources listed in \cite{dcngd}) are shown in Fig.~\ref{fig:reactsys}, for the case of the n-Gd \neb expectation. The total error $\sigma_{\rm{r}}^{i}$ (where $i$ stands for each data point) ranges from 1.75\% (reactors operating at full power) to 1.92\% (one or two reactors not at full power). In a conservative approach, the $\sigma_{\rm{r}}^{i}$ errors are assumed to be fully correlated. 

\section{Background independent oscillation results}

The $R^{\rm{obs}}$ fit is based on a standard $\chi^2$ minimization. Without taking into account the 2-Off data, the $\chi^2$ definition is divided into two different terms: $\chi^2 = \chi^2_{\rm{on}} + \chi^2_{\rm{pull}}$, where $\chi^2_{\rm{on}}$ stands for 2-On and 1-Off reactor data and $\chi^2_{\rm{pull}}$  accounts for the systematic uncertainties. Assuming Gaussian-distributed errors for the data points involving at least one reactor on, $\chi^2_{\rm{on}}$ is built as follows:

\begin{equation}
\label{eq:chi2on}
\chi^2_{\rm{on}} = \sum_i^{\rm{N}} \frac{ \left(R_{i}^{\rm{obs}} - R_{i}^{\rm{exp}}[1+\alpha^{\rm{d}} + k_{i}\alpha^{\rm{r}} + w_{i}\alpha^{\rm{\nu}}]-B\right)^2}{ \sigma_{\rm{stat}}^2},
\end{equation}

\noindent where $N$ stands for the number of bins (6, as shown in Fig.~\ref{fig:expvspth}), and where $\alpha^{\rm{d}}$, $\alpha^{\rm{r}}$ and $\alpha^{\rm{\nu}}$ stand for pulls associated with the detection, reactor-on and residual antineurino systematics, respectively. The weights $k_{i}$ are defined as $\sigma_{\rm{r}}^{i}$/$\sigma_{\rm{r}}$, where $\sigma_{\rm{r}}$=1.75\% stands for the error when the cores operate at full power. The fraction of residual antineurinos, $w_{i}$, in each data point is defined as $w_{i} = R_{i}^{\operatorname{r-\nu}}/R_{i}^{\rm{exp}}$. The term $\chi^2_{\rm{pull}}$ incorporates the penalty terms corresponding to $\sigma_{\rm{r}}$, $\sigma_{\rm{d}}$ and $\sigma_{\rm{\nu}}$:

\begin{equation}
\label{eq:chi2pull}
\chi^2_{\rm{pull}} = \left(\frac{\alpha^{\rm{d}}}{\sigma_{\rm{d}}}\right)^2 + \left( \frac{\alpha^{\rm{r}}}{\sigma_{\rm{r}}} \right)^2 +  \left(\frac{\alpha^{\rm{\nu}}}{\sigma_{\rm{\nu}}}\right)^2.
\end{equation}

\begin{figure}
\begin{center}
\includegraphics[scale=0.4]{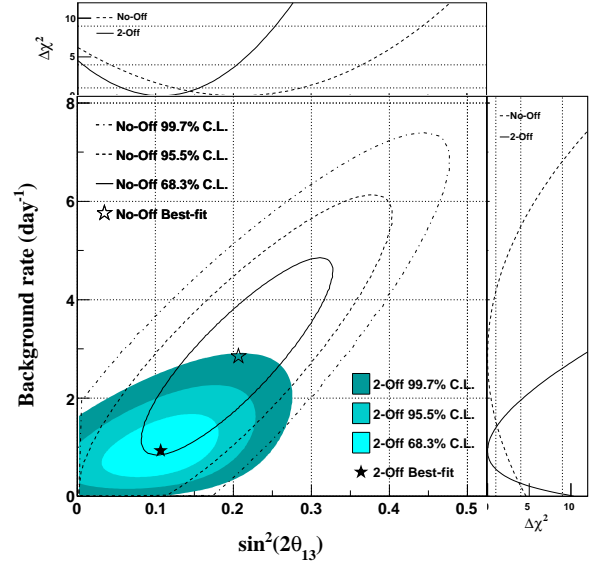}
\caption{RRM ($\sin^2(2\theta_{13})$,$B_{\rm{Gd}}$) fit with n-Gd \neb candidates. Empty (solid) best-fit point and C.L. regions show the results without (with) the 2-Off data sample. 
\label{fig:fit2d}}
\end{center}
\end{figure}

According to this $\chi^2$ definition, a fit to the two free parameters $\sin^2(2\theta_{13})$ and the total background rate $B_{\rm{Gd}}$ is performed with the n-Gd candidates sample. The results are shown in Fig.~\ref{fig:fit2d} with best-fit point (empty star) and C.L. intervals. The best fit values are $\sin^2(2\theta_{13})=0.21\pm0.12$ and $B_{\rm{Gd}}=2.8\pm2.0$ events/day, where the errors correspond to $\Delta\chi^2=2.3$. Although the precision is poor, these results are consistent within 1$\sigma$ with the ones presented in \cite{dcngd}. In particular, the best fit value for the background is consistent with the independent estimate in \cite{dcngd} (1.9$\pm$0.6 events/day) and with the direct measurement obtained from the 2-Off data in \cite{dcoffoff}: $B_{\rm{2Off}}$=0.7$\pm$0.4 events/day (once accidental background is subtracted).

In order to improve the RRM determination of $\sin^2(2\theta_{13})$, the 2-Off data can be incorporated into the fit as an additional data point for $P_{\rm{th}}=0$ MW. The $\chi^2$ is built then as $\chi^2 = \chi^2_{\rm{on}} + \chi^2_{\rm{off}} + \chi^2_{\rm{pull}}$. Due to the low n-Gd statistics in the 2-Off reactor period, the corresponding error in $R^{\rm{obs}}$ is considered to be Poisson-distributed. As a consequence, $\chi^2_{\rm{off}}$ is defined as a binned Poisson likelihood following a $\chi^2$ distribution:

\begin{eqnarray}
\label{eq:chi2off}
\nonumber\chi^2_{\rm{off}}  & = &2 \Big( N^{\rm{obs}}\textrm{ln} \frac{N^{\rm{obs}}}{B+N^{\rm{exp}}[1+\alpha^{\rm{d}}+\alpha^{\rm{\nu}}]} \\
           &  & +B+N^{\rm{exp}}[1+\alpha^{\rm{d}}+\alpha^{\rm{\nu}}] - N^{\rm{obs}}\Big),
\end{eqnarray}

\noindent where $N^{\rm{obs}}=R^{\rm{obs}}\cdot T_{\rm{off}}$ and $N^{\rm{exp}}=R^{\operatorname{r-\nu}}\cdot T_{\rm{off}}$; $T_{\rm{off}}$ the live time of the 2-Off data sample. The results of the ($\sin^{2}(2\theta_{13})$,B) fit including the 2-Off data are presented in Fig.~\ref{fig:fit2d} with solid best fit point and C.L. intervals.  The best fit values are $\sin^2(2\theta_{13})=0.107\pm0.074$ and $B_{\rm{Gd}}=0.9\pm0.6$ events/day.

\begin{figure}
\begin{center}
\includegraphics[scale=0.4]{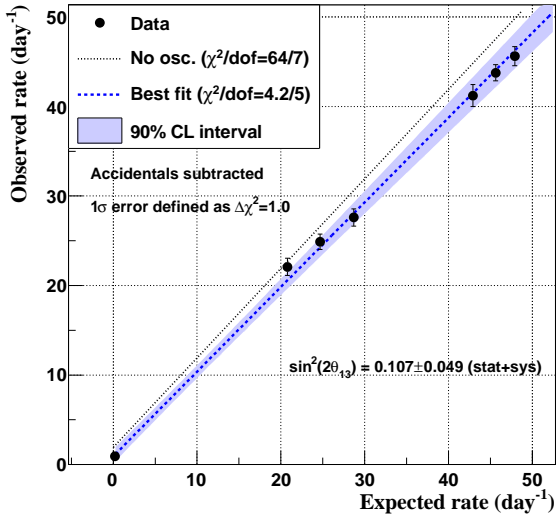}
\caption{RRM fit with n-Gd \neb candidates including 2-Off data. The null oscillation hypothesis assuming the background estimates obtained in \cite{dcngd} is also shown for comparison purposes. 
\label{fig:fit1}}
\end{center}
\end{figure}

As the 2-Off data provide the most precise determination of the total background rate in a model-independent way, the introduction of this sample (or equivalently the value of $B_{\rm{2Off}}$) in the RRM fit provides a direct constraint to $B$. Therefore, hereafter we consider $\theta_{13}$ to be the only free parameter in the fit, while $B$ is treated as a nuisance parameter. Therefore, the best fit error on $\theta_{13}$ corresponds to $\Delta\chi^2=1$. The outcome of the corresponding fit using the n-Gd sample can be seen in Fig.~\ref{fig:fit1}. The best fit value of $\sin^{2}(2\theta_{13})$ is now 0.107$\pm$0.049,  with a $\chi^2$/dof of 4.2/5. The value of $\theta_{13}$ is in good agreement with the result of \cite{dcngd} ($\sin^{2}(2\theta_{13})$=0.109$\pm$0.039), while the error is slightly larger due to the fact that the RRM analysis does not incorporate energy spectrum information. The RRM fit does not change the measurement of the total background rate provided by the 2-Off data significantly, as the best fit estimate of $B_{\rm{Gd}}$ is 0.9$\pm$0.4 events/day.      

While the best fit value of the total background rate depends on the antineurino candidate selection cuts, the best fit of $\theta_{13}$ must be independent of these cuts. In order to cross-check the above results, the RRM  analysis has also been performed for a different set of selection cuts: those applied in the first Double Chooz oscillation analysis \cite{dcngd1st}. This selection does not make use of the muon outer veto (OV) and does not apply a showering muon veto. Therefore, the number of correlated background events in the \neb candidates sample is increased (according to the estimates, by 1.3 events/day). In this case, the input value for the background rate provided by the 2-Off data is $B_{\rm{2Off}}=2.4\pm0.6$ events/day \cite{dcoffoff}. The fit yields sin$^2(2\theta_{13})$=0.120$\pm$0.053, which is fully consistent with the above results, while the background rate is not significantly modified either in this case ($B_{\rm{Gd}}$=2.6$\pm$0.6 events/day).


As shown in \cite{dcnh}, the precision of the oscillation analysis based on n-H captures is not as good as the n-Gd one due to the larger systematic uncertainties and the larger accidental contamination. This applies also to the RRM analysis. The n-H fit yields $\sin^2(2\theta_{13})$=0.091$\pm$0.078 ($B_{\rm{2Off}}$=10.8$\pm$3.4 events/day, $B_{\rm{H}}$=8.7$\pm$2.5 events/day) with $\chi^2/dof$=4.8/5, consistent with the results in \cite{dcnh}($\sin^2(2\theta_{13})$=0.097$\pm$0.048). The n-H candidates can be fitted together with the n-Gd ones in order to increase the precision of the analysis and to test the consistency of both selections.  In order to perform a global fit, a combined $\chi^{2}$ is built from the sum of the Gd and H ones:

\begin{equation}
\label{eq:chi2}
\chi^2 = \chi^2_{\rm{Gd}} + \chi^2_{\rm{H}} +\chi^2_{\rm{pull}}.
\end{equation}

While $\sigma_{\rm{r}}$ and $\sigma_{\rm{\nu}}$  are fully correlated between the n-Gd and n-H candidates samples (they do not depend on selection cuts, but on reactor parameters), there is a partial correlation ($\rho$) in the detection efficiency uncertainty, which has been estimated to be at the level of 9\%. This overall factor comes from correlated and anti-correlated contributions. The correlated contributions are due to the spill-in/out events (IBD events in which the prompt and the delayed signal do not occur in the same detection volume, as defined in \cite{dcngd}) and the number of protons in the detection volumes. The anti-correlated contribution is due to the uncertainty in the fraction of neutron captures in Gd and H. From this $\rho$ value, one can decompose $\sigma_{\rm{d}}$ into uncorrelated ($\sigma^{\rm{d}}_{\operatorname{Gd-u}}=0.91\%$ and $\sigma^{\rm{d}}_{\operatorname{H-u}}=1.43\%$) and correlated contributions ($\sigma^{\rm{d}}_{\rm{c}}=0.38\%$) for the n-Gd and n-H data. The pull $\alpha^{\rm{d}}$ in Eq.~\ref{eq:chi2pull} is now divided into three terms accounting for the correlated and uncorrelated parts of the detection error: $\alpha^{\rm{d}}_{\operatorname{Gd-u}}$, $\alpha^{\rm{d}}_{\operatorname{H-u}}$ and $\alpha^{\rm{d}}_{\rm{c}}$. Accordingly, $\chi^2_{\rm{pull}}$ is defined as:

\begin{eqnarray}
\label{eq:chi2pen}
\nonumber\chi^2_{\rm{pull}} & = &  \left(\frac{\alpha^{\rm{d}}_{\operatorname{Gd-u}}}{\sigma^{\rm{d}}_{\operatorname{Gd-u}}}\right)^2 + \left(\frac{\alpha^{\rm{d}}_{\operatorname{H-u}}}{\sigma^{\rm{d}}_{\operatorname{H-u}}}\right)^2+ \left(\frac{\alpha^{\rm{d}}_{c}}{\sigma^{\rm{d}}_{\rm{c}}}\right)^2  \\
& & + \left(\frac{\alpha^{\rm{r}}}{\sigma^{\rm{r}}}\right)^2 +  \left(\frac{\alpha^{\rm{\nu}}}{\sigma_{\rm{\nu}}}\right)^2.
\end{eqnarray}

\begin{figure}
\begin{center}
\includegraphics[scale=0.4]{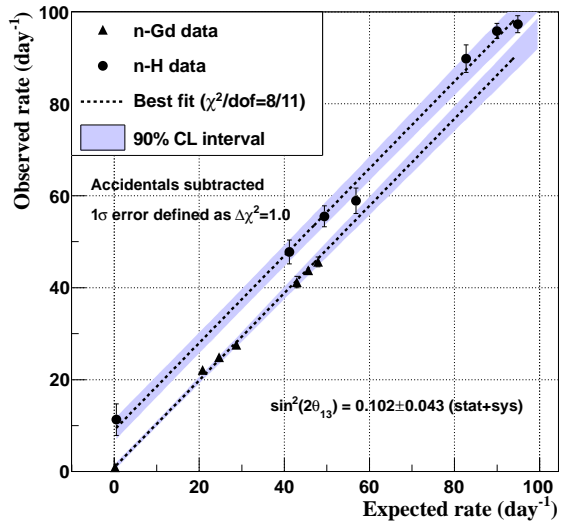}
\caption{ RRM combined fit using n-Gd and n-H \neb candidates.
\label{fig:fitResults2}}
\end{center}
\end{figure}

The combined Gd-H RRM fit is shown in Fig.~\ref{fig:fitResults2}. The best fit value of the mixing angle is $\sin^2(2\theta_{13})=0.102\pm$0.028(stat.)$\pm$0.033(syst.), for $\chi^2/dof=8.0/11$. This value is consistent within 1$\sigma$ with respect to the single n-Gd and n-H results, while the precision is slightly improved. The relative error on $\sin^2(2\theta_{13})$ goes from 46\% to 42\%. As in the previous results, the output values of the total background rates are consistent with the input values: $B_{\rm{Gd}}=0.9\pm0.4$ events/day and $B_{\rm{H}}=9.0\pm1.5$ events/day. The impact of the correlated part in $\sigma^{\rm{d}}$ has been proven to be negligible by performing a fit assuming no correlation.

\section{Comparison of Double Chooz $\theta_{13}$ results}

\begin{figure}
\begin{center}
\includegraphics[scale=0.4]{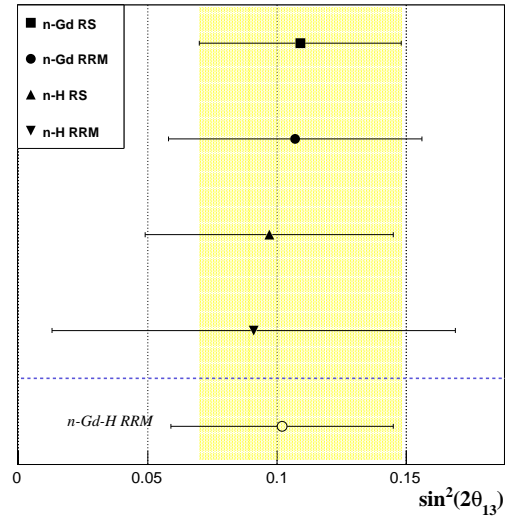}
\caption{ Summary of published Double Chooz results on $\theta_{13}$: the n-Gd \cite{dcngd} and n-H \cite{dcnh} rate plus shape (RS) results, and the n-Gd and n-H RRM ones. For comparison purposes, the combined Gd-H RRM result is also shown. The shaded region shows the 68\% C.L. interval of the n-Gd RS fit.
\label{fig:Results}}
\end{center}
\end{figure}

Including this novel RRM analysis, Double Chooz has released four different $\theta_{13}$ analysis results. These results are obtained as follows: 1) with n-Gd candidates in \cite{dcngd}, 2) with n-H candidates in \cite{dcnh}, 3) with n-Gd candidates and the RRM analysis, and 4) with n-H candidates and the RRM analysis. Beyond the common detection and reactor-related systematics, these four analyses rely on two different candidate samples (n-H and n-Gd), and two different analysis techniques (rate+shape fit with background inputs and RRM). The four $\sin^2(\theta_{13})$ values obtained are presented in Fig.~\ref{fig:Results}, as well as the combined Gd-H RRM result. All the values are consistent within 1$\sigma$ with respect to the most precise result, which is provided by the rate+shape fit.

\section{Summary and conclusions}

While the oscillation results published by the Double Chooz collaboration in \cite{dcngd,dcnh,dcngd1st} rely on a background model derived from reactor-on data, the RRM analysis is a background model independent approach. Both $\theta_{13}$ and the total background rate are derived without model assumptions on the background by a global fit to the observed antineutrino rate as a function of the non-oscillated expected rate for different reactor power conditions. Although the RRM fit with only reactor-on data does not achieve a competitive precision on $\theta_{13}$, it provides an independent determination of the total background rate. This rate is consistent with the Double Chooz background model and with the measurement of the total background from the 7.53 days of reactor-off data \cite{dcoffoff}. As this 2-Off sample provides the most precise determination of the total background rate in a model independent way, it is introduced in the RRM analysis in order to improve the results on $\theta_{13}$, which remains as the only free parameter in the fit. The best fit value of $\sin^{2}(2\theta_{13})$=0.107$\pm$0.049 is found by analyzing the n-Gd \neb candidates. Finally, the precision on $\theta_{13}$ is further improved by combining the n-Gd and n-H \neb samples:  $\sin^2(2\theta_{13})=0.102\pm$0.028(stat.)$\pm$0.033(syst.). The outcome of the RRM fit is consistent within 1$\sigma$ with the already published results for $\theta_{13}$, yielding a competitive precision. Beyond the cross-check of the background estimates in the Double Chooz oscillation analyses, the RRM analysis provides, for the first time, a background model independent determination of the $\theta_{13}$ mixing angle.

\begin{acknowledgments}
We thank the French electricity company EDF; the
European fund FEDER; the R\'egion de Champagne Ardenne;
the D\'epartement des Ardennes; and the Communaut\'e des Communes Ardennes Rives de Meuse. We
acknowledge the support of the CEA, CNRS/IN2P3, the
computer center CCIN2P3, and LabEx UnivEarthS in
France; the Ministry of Education, Culture, Sports, Science
and Technology of Japan (MEXT) and the Japan
Society for the Promotion of Science (JSPS); the Department
of Energy and the National Science Foundation of
the United States; the Ministerio de Ciencia e Innovaci\'on
(MICINN) of Spain; the Max Planck Gesellschaft,
and the Deutsche Forschungsgemeinschaft DFG (SBH
WI 2152), the Transregional Collaborative Research Center TR27, the 
excellence cluster ``Origin 
and Structure
of the Universe'', and the Maier-Leibnitz-Laboratorium
Garching in Germany; the Russian Academy of
Science, the Kurchatov Institute and RFBR (the Russian
Foundation for Basic Research); the Brazilian Ministry of
Science, Technology and Innovation (MCTI), the Financiadora
de Estudos e Projetos (FINEP), the Conselho
Nacional de Desenvolvimento Cient\'ifico e Tecnol\'ogico
(CNPq), the S\~ao Paulo Research Foundation (FAPESP),
and the Brazilian Network for High Energy Physics (RENAFAE)
in Brazil.
\end{acknowledgments}


\begin{thebibliography}{99}

\bibitem{dcngd}
  Y.~Abe {\it et al.}  [Double Chooz Collaboration],
  Phys.\ Rev.\ D {\bf 86} (2012) 052008
  [arXiv:1207.6632 [hep-ex]].

\bibitem{db}
F.P. ~An {\it et al.} [Daya Bay Collaboration],
Phys.\ Rev.\ Lett. {\bf 108} (2012) 171803
[arXiv:1203.1669 [hep-ex]].


\bibitem{reno}
J.K. ~Ahn {\it et al.} [Reno Collaboration],
Phys.\ Rev.\ Lett. {\bf 108} (2012) 191802
[arXiv:1204.0626 [hep-ex]].


\bibitem{dcoffoff}
  Y.~Abe {\it et al.}  [Double Chooz Collaboration],
  Phys.\ Rev.\ D {\bf 87} (2013) 011102
  [arXiv:1210.3748 [hep-ex]].
  
\bibitem{minos} P. Adamson {\it et al.} [MINOS Collaboration], Phys. Rev. Lett. 106, 181801 (2011).

\bibitem{dcnh}
  Y.~Abe {\it et al.}  [Double Chooz Collaboration],
  Phys.\ Lett.\ B {\bf 723} (2013) 66
  [arXiv:1301.2948 [hep-ex]].

\bibitem{dcngd1st}
  Y.~Abe {\it et al.}  [Double Chooz Collaboration],
  Phys.\ Rev.\ Lett.\  {\bf 108} (2012) 131801
  [arXiv:1112.6353 [hep-ex]].

\end{thebibliography}
\end{document}